\documentclass[12pt,preprint]{aastex}
\usepackage{graphicx}
\usepackage{amssymb,amsmath}

\pdfoutput=1

\newcommand\beq{\begin{equation}}
\newcommand\eeq{\end{equation}}
\newcommand\lsim{\mathrel{\rlap{\lower4pt\hbox{\hskip1pt$\sim$}}
        \raise1pt\hbox{$<$}}}
\newcommand\gsim{\mathrel{\rlap{\lower4pt\hbox{\hskip1pt$\sim$}}
        \raise1pt\hbox{$>$}}}

\newcommand{\degrees}{\mbox{$^\mathrm{o}$}}
\newcommand{\ms}{m~s$^{-1}$}

\begin{document}
\title{Atmospheric Circulation and Composition of GJ1214b}

\author{Kristen
  Menou\altaffilmark{1}}
 
\altaffiltext{1}{Department of Astronomy, Columbia University, 550
  West 120th Street, New York, NY 10027}

\begin{abstract}
The exoplanet GJ1214b presents an interesting example of compositional
degeneracy for low-mass planets. Its atmosphere may be composed of
water, super-solar or solar metallicity material. We present
atmospheric circulation models of GJ1214b for these three
compositions, with explicit grey radiative transfer and an optional
treatment of MHD bottom drag. All models develop strong, superrotating
zonal winds ($\sim 1$-$2$ km/s). The degree of eastward heat
advection, which can be inferred from secondary eclipse and thermal
phase curve measurements, varies greatly between the models. These
differences are understood as resulting from variations in the
radiative times at the thermal photosphere, caused by separate
molecular weight and opacity effects. Our GJ1214b models illustrate
how atmospheric circulation can be used as a probe of composition for
similar tidally-locked exoplanets in the mini-Neptune/waterworld
class.
\end{abstract}

\section{Introduction}

The exoplanet GJ1214b has a mass of $6.5 M_\earth$ and a radius of
$2.65 R_\earth$ (Charbonneau et al. 2009). This mass-radius
combination does not permit a unique inference of the bulk composition
of the planet, because of significant compositional degeneracies in
the structural models (e.g. Fortney et al. 2007; Adams et
al. 2008). Possibilities include a substantial envelope of water or
hydrogen-dominated gas (perhaps enriched in heavy elements), above a
massive solid/icy core (Rogers \& Seager 2010; Nettelmann et
al. 2011). We refer to such low mass exoplanets with deep envelopes of
H or H$_2$O as mini-Neptunes/waterworlds.

In a compositionally degenerate situation like that of GJ1214b, a
promising method to constrain the atmospheric composition is to obtain
a transmission spectrum of the planet as it transits its parent
star. Indeed, the depth of atmospheric absorption features during
transit is a direct measure of the atmosphere's mean molecular weight
(Miller-Ricci et al. 2009; Miller-Ricci \& Fortney 2010). Recently,
such measurements have been performed for GJ1214b by several groups,
at different wavelengths. Several reported transit spectra suggest a
significantly metal-enriched or water atmosphere for GJ1214b (Bean et
al. 2010, 2011; Desert et al. 2011; Crossfield et al. 2011), with one
conflicting suggestion of a metal-poor atmospheric composition (Croll
et al. 2011).

Here, we propose a complimentary probe of atmospheric composition for
tidally-locked mini-Neptunes/waterworlds based on their specific
regime of atmospheric circulation. Using GJ1214b as an example, we
show that variations in atmospheric mean-molecular weight and
opacities can have a strong enough effect on the degree of eastward
heat advection at the thermal photosphere that secondary eclipses and
thermal phase curves may also be used to consolidate our knowledge of
the atmospheric composition of these planets. In \S2, we describe our
setup for modeling atmospheric circulation on GJ1214b, using three
different compositions. Our model results and their interpretation are
presented in \S3, before we conclude in \S4.

\section{Models}

To model the atmospheric circulation of GJ1214b, we use the
Intermediate General Circulation Model (IGCM; Hoskins \& Simmons
1975).  It is a well-tested and accurate solver of the primitive
equations of meteorology, which are satisfied by a shallow atmosphere
of ideal gas in hydrostatic balance on a rotating planet.  We use IGCM
version 3, which contains several major improvements over earlier
versions (Forster et al. 2000).  The pseudo-spectral, semi-implicit
dynamical component is described and tested in Menou \& Rauscher
(2009).  Our implementation of IGCM3 for the study of gaseous
exoplanets is described in Rauscher \& Menou (2011b).

We assume that GJ1214b has a circular orbit and that the planet's
rotation is tidally synchronized with its orbit.\footnote{We refer to
  the rotational (and orbital) period of the planet as a planet day
  ($\simeq 1.58$~Earth days).} This leads to steady hemispheric
forcing of the permanently irradiated dayside, with a substellar flux
of $2.15 \times 10^4$~W~m$^{-2}$ for GJ1214b. The two-stream radiative
transfer scheme employed, which was adapted from the non-grey version
of Forster et al. (2000), treats separately the shortwave (stellar)
and longwave (thermal) radiation, with two different grey
opacities. It is the same scheme as the one extensively tested in
Rauscher \& Menou (2011b), except that we do not implement a switch to
the diffusion approximation in the model deepest layers here.

For numerical stability and to model small-scale dissipation (e.g.,
Thrastarson \& Cho 2011), hyperdissipation is included. It acts upon
the vertical component of the flow relative vorticity, the flow
divergence, and the temperature fields with an iterated Laplacian of
order 8 and a hyperdissipation coefficient $\nu_{\mathrm{diss}} =6.9
\times 10^{43}$~m$^8$~s$^{-1}$ chosen so that the smallest resolved
structures are diffused in $0.25$ planet day.

A dry convective adjustment scheme is used to mix entropy at each
timestep in all the atmospheric columns found to be convectively
unstable. In practice, given the stably-stratified nature of the
vertical profiles emerging in our models, convective adjustment is
typically used only early in the runs, when the flow adjusts to the
radiative forcing from the prescribed initial conditions.  The
atmosphere is started at rest. The initial temperature field is set as
quasi-isothermal at $T =920 $~K and subjected to spatially uniform
1D-averaged irradiation for 0.5 planet days before the hemispheric
forcing is imposed. (Other comparable temperature initial conditions
yield similar model results.) To break initial symmetry, we also
include noise in the form of small amplitude random perturbations in
the surface pressure field.

{\bf Numerical Resolution and Convergence} -- All six models presented
here were run at T31L30 resolution, which corresponds to slightly less
than 4\degrees~in latitude and longitude, with 30 logarithmically
spaced vertical levels.  The vertical domain extends from 0.1 mbar to
10 bar, so that our bottom boundary is above the radiative-convective
boundary at $\gsim 10$~bar for the compositions of interest
(Miller-Ricci \& Fortney 2010). We also ran $\sim 50$ additional
models at lower resolutions (T10L20 and T21L20) to evaluate numerical
convergence and to survey the parameter space of circulations on
GJ1214b, with variations in opacities, mean molecular weight,
intrinsic heat flux (up to $1\%$ of the irradiation flux) and bottom
drag. These lower resolution results are generally consistent with the
higher resolution versions presented in Figs~1-4, with minor
quantitative differences.

{\bf Composition, Opacities and Mean Molecular Weights} -- The
composition of the atmosphere of GJ1214b is presently
unknown. Possibilities include a composition that is H-dominated, with
solar or super-solar composition, CO$_2$-dominated or H$_2$O-dominated
(Rogers \& Seager 2010). Since our main interest here is to explore
the interplay between atmospheric circulation and composition, we
choose to focus on three particular compositions: water, $\times 30$
supersolar and solar composition. These three compositions allow us to
separate rather well the effects of molecular weight and opacity
variations on the atmospheric circulation regime, as described
below. For convenience, we refer to models with such compositions as
Water, Supersol and Solar.

The 1D radiative-convective models of GJ1214b by Miller-Ricci \&
Fortney (2010) show that temperature-pressure profiles, and therefore
opacities, are quite similar for a water and a $\times 30$ supersolar
composition. From the point of view of atmospheric circulation,
different behaviors between such atmospheres will thus be mostly
caused by differences in mean molecular weight. By contrast, the Solar
model has a molecular weight only moderately smaller than that of the
Supersol model, but substantially weaker opacities. Therefore, our
choice of three compositions emphasizes differences in mean molecular
weights (Water vs. Supersol) and in opacities (Supersol
vs. Solar). Using mean molecular weights $\mu \simeq 18$, $3.3$ and
$2.2$ for the Water, Supersol and Solar models, we deduce the gas
constant values listed as ${\cal R}$ in Table~1. We assume the same
value of $\kappa = {\cal R}/C_p$ in all models, where $C_p$ is the
specific heat at constant pressure (see Table~1).

We choose the grey opacity coefficients in our models so as to match
as closely as possible the temperature-pressure profiles of
Miller-Ricci \& Fortney (2010) for the Water, Supersol ($\times 30$)
and Solar compositions, using the simple grey radiative solutions of
Guillot (2010) as a comparison tool.\footnote{Our two-stream radiative
  scheme has been successfully benchmarked against Guillot's solutions
  (Rauscher \& Menou 2011b).} We find reasonable matches for the
values of the visible and thermal opacity coefficients listed in
Table~1.  We pay close attention to the location of
the shortwave and longwave photospheres, indicated by turning points
in the profiles, using the same dayside-averaging as Miller-Ricci \&
Fortney (2010). We neglect the planet's intrinsic heat flux, which
does not have a major impact at the atmospheric levels of
interest. While our matches with such a grey model can only be
approximate, we do not expect minor discrepancies in
temperature-pressure profiles to affect our main conclusions. Indeed,
our results largely rely on the simple effect that strong molecular
weight and opacity variations have on the radiative times at the
thermal photosphere (see \S\ref{sec:results}).

{\bf MHD Bottom Drag} -- An outstanding question for the modeling of
atmospheric circulation on planets with deep atmospheres is the nature
of drag mechanisms limiting the atmospheric wind speeds. A promising
mechanism for hot Jupiters with radiative equilibrium temperatures
$T_{\rm eq} \gsim 1000$~K is the magnetic drag that results from
induction by weakly-ionized zonal winds flowing through the poloidal
planetary magnetic field (Perna et al. 2010; Rauscher \& Menou
2011a,b; Menou 2011). A related mechanism has been proposed to limit
wind speeds in the deep layers of the atmospheres of Solar System
giant planets, from interactions with the interior adiabat in presumed
solid-body rotation (Schneider \& Liu 2009; see also Liu et
al. 2008). Altogether, these arguments suggest that MHD drag may also
be acting in the deep layers of the atmosphere of GJ1214b, possibly in
the vicinity of the radiative-convective boundary, where atmospheric
temperatures quickly rise above $1000$~K (Miller-Ricci \& Fortney
2010).

In the absence of a detailed drag model, we include MHD bottom drag in
some of our models by applying linear Rayleigh drag in the two deepest
layers. We adopt drag times of $40$ and $20$ planet days for the
second-to-deepest and deepest model layers, respectively. This may be
interpreted as including a weak boundary layer in our models, to
represent interaction with the convective interior maintained in near
solid-body rotation by the effects of MHD rigidity (Liu et
al. 2008). Given the close proximity of our deepest model layers to
the radiative-convective boundary, we do not include any latitudinal
dependence of the drag, as may be geometrically more appropriate for
Solar System gaseous giant planets with MHD coupling only at
significant depth below the atmosphere (Schneider \& Liu 2009).

We have verified that our main results do not depend on specifics of
the MHD bottom drag treatment (one- vs. two-layer drag, stronger
vs. weaker drag), although some detailed aspects of the circulation
regime can (see \S\ref{sec:results}). This robustness is illustrated
in Fig.~4, where models with and without bottom drag show consistent
results. Table~1 lists the values of all the other important physical
parameters adopted in our models of GJ1214b. To summarize, all six
models are identical except for different values of the gas constant
(mean molecular weight) and the opacity coefficients, with and without
bottom drag.

\section{Results} \label{sec:results}

All model results are shown after long numerical integration of $5000$
planet days ($= 7800$ Earth days). We find that steady-state is
established well earlier than this for all the quantities shown in
Figs~1-4. Spin-up is typically achieved after a few thousand planet
days for the deepest model layers.

Figure~1 shows temperature-pressure\footnote{We use pressure as the
  vertical coordinate in our figures, rather than the model coordinate
  system, $\sigma = p/p_s$, because horizontal variations in surface
  pressures, $p_s$, are small enough in the models that the
  distinction between $p$ and $\sigma$ is unimportant.} profiles in
the Water model with bottom drag, for six different columns around the
planet. Similar profiles are obtained in the Supersol model, with
identical opacities. The profiles in the Solar model have broadly
similar shapes but they are systematically shifted down in pressure,
as a result of reduced opacities (see also Fig.~1 of Miller-Ricci \&
Fortney 2010 for such a shift). In all models, the poles are
significantly colder than the equator. This feature, which can be
attributed to limited meridional heat transport relative to the zonal
transport, is also clearly apparent in the thermal flux maps (Fig.~3).

Figure~2 shows contour plots of the zonal average of the zonal wind
velocity for the three models with drag, as a function of latitude and
pressure in the atmosphere. A broad superrotating equatorial jet with
velocities $\sim 1$-$2$~km~s$^{-1}$ is a common feature of all three
models. Winds are also eastward (positive) almost everywhere in these
deep atmospheres, with noticeably stronger winds at high latitude in
the Supersol model, and particularly the Solar model. Westward
(negative) winds are weak and only present in the deepest layers,
where bottom drag is applied.

Our results are consistent with the theory of Showman \& Polvani
(2011), which points to the formation of superroting equatorial jets
under a broad range of atmospheric conditions for tidally-locked
planets subject to dayside hemispheric forcing.\footnote{Small
  rotational offsets from tidal-locking would yield similar resuls,
  but it is unclear what circulation regime and thermal signatures
  would emerge in the case of substantial deviations from
  tidal-locking.} We estimate similar Rossby numbers, $ {\rm Ro} \sim
1.5$, and scaled Rhines lengths, $L_\beta/R_p \sim 3.5$, in all our
models, for a 1~km~s$^{-1}$ wind scale (see Showman et al. 2010 for a
review of basic atmospheric scales). By contrast, the scaled Rossby
deformation radius is $L_D/R_p \sim 0.7$ in the Water model, $\sim
1.5$ in the Supersol model and $\sim 1.8$ in the Solar model. Larger
Rossby deformation radii in the Supersol and Solar models may be
responsible for their broader patterns of eastward zonal winds, as
seen in Fig.~2 (see also Showman \& Polvani 2011).

In the three models shown in Fig.~2, the atmosphere must reach
momentum balance with the drag applied at the bottom, which represents
an interaction with an effectively infinite reservoir of momentum in
the deep interior. By contrast, atmospheric angular momentum is
globally conserved in the three models without drag (not shown). This
important distinction could in principle have a strong impact on the
atmospheric circulation regime (e.g., Rauscher \& Menou 2010). We find
no such major difference in our Supersol and Solar models, but there
is a qualitative difference in circulation regime for the Water model
without drag (not shown) relative the one with drag (top panel,
Fig~2). In the drag-free model, which shows some signs of additional
dynamical activity, north-south symmetry is broken. Winds in the
northern hemisphere, beyond the superrotating equatorial jet, are
markedly westward (negative), throughout the depth of the
atmosphere. While this difference between the Water models with and
without drag is clearly important, we have not explored these
dynamical issues any further. Indeed, the key dynamical feature that
is common to all our models is the superrotating equatorial jet, which
predominantly determines the thermal signatures of GJ1214b, whether or
not bottom drag is applied (see Fig.~4).

Figure~3 shows maps of outgoing thermal flux for the same three models
with bottom drag as shown in Fig.~2. Since these maps are centered on
the substellar point, they reveal a clearly incremental amount of
eastward heat advection (and re-radiation), from the Water model (top)
to the Supersol model (middle) and finally the Solar model
(bottom). By contrast, heat advection towards the poles remains
limited in all three models.

Figure~4 confirms these trends by showing the thermal phase curves
expected for an observer located at a $90\deg$ inclination with
respect to the planet's rotational axis. These phase curves are
calculated by integrating the flux from each visible atmospheric
column at each location along the planet's orbit, accounting for the
proper geometry of emission (see Rauscher \& Menou 2011b for
details). Differences in the thermal phase curves of models with and
without bottom drag (solid vs. dashed lines) are very small or
negligible. The phase curve contrast, measured as the ratio of the
maximum to minimum thermal flux, decreases from $1.9$ in the Water
model to $1.4$ in the Supersol model and $1.09$ in the Solar
model. The angular offset at peak emission, relative to the time of
secondary eclipse at phase 0.5, is of order $ 50\deg$ in the Water and
Supersol models, and $ 100 \deg$ in the Solar model.  Adopting the
thermal flux at phase 0.5 as a measure of secondary eclipse depth, we
deduce an eclipsed flux of order $ 7000$~W~m$^{-2}$ for the Water
model, $ 6200$~W~m$^{-2}$ for the Supersol model and $
5650$~W~m$^{-2}$ for the Solar model. Even for the Solar model with
strong zonal transport, the eclipsed flux is above the
$5375$~W~m$^{-2}$ expected for perfect heat redistribution, because
the polar regions remain cooler than the equator (see Fig.~3).

The differences in thermal curves shown in Fig.~4 can easily be
understood as resulting from molecular weight and opacity effects. In
all our model, the timescales to advect heat away from the substellar
point and around the planet's equator at the thermal photosphere are
comparable, with $\tau_{\rm adv} \sim R_p/U $ and wind velocities $U
\sim 1$-$2$~km/s at the thermal photosphere. By contrast, radiative
times at the thermal photosphere vary substantially from one model to
the next. Since $\tau_{\rm rad} \sim C_pP/(g \sigma T^3)$, we estimate
that $\tau_{\rm rad}$ is about $5.5$ times longer in the Supersol
model than in the Water model, in proportion to their mean molecular
weights (or gas constants, ${\cal R}$). The difference in mean
molecular weights between the Supersol and Solar models is a modest
factor $\sim 1.5$, but the significantly weaker opacities in the Solar
model have a large effect on the location of the thermal
photosphere. A factor 5 reduction in opacities results in a thermal
photosphere at 5 times greater pressure in the Solar model. Factoring
both opacity and molecular weight effects, this implies that
$\tau_{\rm rad}$ is approximately $7.5$ times longer in the Solar
model than it is in the Supersolar one. The substantial decrease of
the ratio $\tau_{\rm adv}/ \tau_{\rm rad}$ in the models shown in
Figs.~3 and~4, from top to bottom, combined with the prevalence of
superrotating equatorial jets in all models, leads to an increasing
amount of eastward heat advection (e.g. Cowan \& Agol 2011; Menou
2011), which is largely responsible for the differences in phase
curves shown in Fig.~4.

\section{Conclusion}

Our results for GJ1214b are interesting in and of themselves because
there is significant observational potential for obtaining robust
constraints on the thermal properties of this particular exoplanet. Of
course, our models of the atmospheric circulation on GJ1214b, with
simple grey radiative transfer and bottom drag treatment, remain
idealized. Various additional considerations could complicate our
attempts to understand the regime of circulation on GJ1214b, such as
the presence of clouds/hazes in the atmosphere (Miller-Ricci et
al. 2011) or the possibility that the planet maintains a finite
eccentricity (Charbonneau et al. 2009).  This can be addressed by
building models of increasing complexity as more observational
constraints become available.

Although applied specifically to GJ1214b, our main arguments are
simple enough that they may retain validity for other similar
tidally-locked mini-Neptunes/waterworlds, such as HD97658b (Henry et
al. 2011) and perhaps Kepler-11b and~11c (Lissauer et al. 2011).  To
the extent that superrotating equatorial jets are robust features of
the atmospheric circulation regime on such planets, as suggested by
our results and recent work on hot Jupiters (Showman \& Polvani 2011)
and hot Neptunes (Lewis et al. 2010), one can expect differences in
mean-molecular weight and opacities to result in distinct thermal
signatures of a similar character as the ones shown in our Fig.~3
and~4. Such considerations, combined with planet-specific atmospheric
models like the ones presented here, should help us constrain the
atmospheric composition of tidally-locked mini-Neptunes/waterworlds.

\acknowledgments

The author thanks Emily Rauscher for providing some of the routines
used in this study. This work was supported in part by NASA grant PATM
NNX11AD65G.

\newpage

\begin{deluxetable}{lccc} 
\tablewidth{0pt}
\tablecaption{Model Parameters}
\tablehead{
\colhead{Parameters}  & \multicolumn{3}{c}{Model} \\
\colhead{~}  & \colhead{Water}  & \colhead{SuperSol}     & \colhead{Solar} 
}
\startdata
$g$ (gravitational acceleration [m s$^{-2}$])&  8.93   &   8.93&  8.93  \\
$\Omega_p$  (planetary rotation rate [rad s$^{-1}$]) &  $4.615 \times 10^{-5}$   &    $4.615 \times 10^{-5}$ &  $4.615 \times 10^{-5}$   \\
$R_p$  (planetary radius  [m]) &  $1.7 \times 10^7$   &    $1.7 \times 10^7$&  $1.7 \times 10^7$     \\
${\cal R}$  (perfect gas constant [MKS]) &  483   &  2519  &    3779    \\
$\kappa$  ($={\cal R}/c_p$ ) &  0.286   &   0.286 &  0.286     \\ 
\\ \hline\\
$k_{\rm th}$ (thermal opacity coeff. [cm$^2$~g$^{-1}$]) & $0.1$ & $0.1$  &$2 \times 10^{-2}$\\
$k_{\rm vis}$ (visible opacity coeff. [cm$^2$~g$^{-1}$]) & $4 \times 10^{-3}$ & $4 \times 10^{-3}$  &  $8 \times 10^{-4}$\\

\enddata
\label{tab:one}
\end{deluxetable}

\clearpage

\begin{figure*}[l]
\centering \includegraphics[scale=0.6]{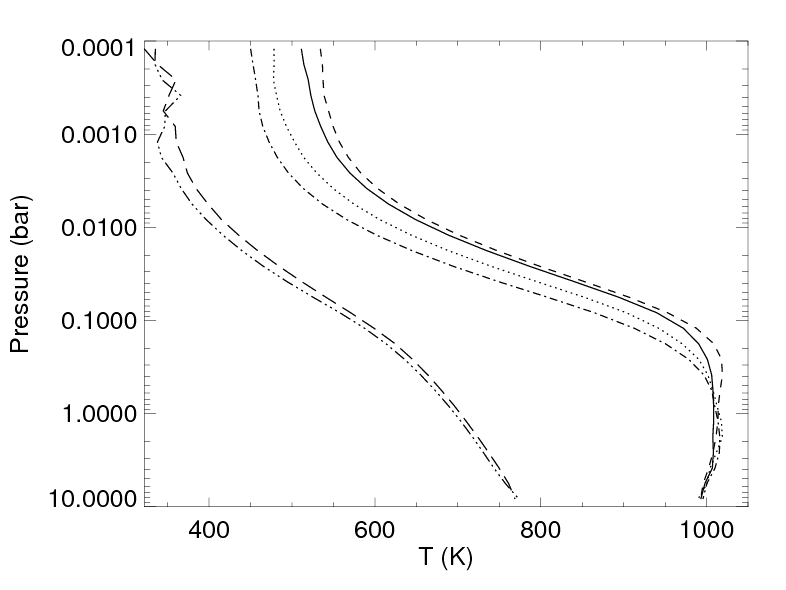}
\caption{Examples of temperature-pressure profiles in the Water model
  with bottom drag, at six different locations around the planet
  (substellar point = {solid} line, antistellar point = {dotted} line,
  equator at east terminator = {dashed} line, equator at west
  terminator = {dot-dashed} line, north pole = {triple dot-dashed}
  line, south pole = {long dashed} line). Polar regions are
  significantly colder than the equatorial regions. The thick solid
  line shows the dayside-averaged profile obtained from Guillot's
  (2010) radiative solution for identical opacities. The use of 1D
  averaged models to describe the transmission spectroscopic
  signatures of GJ1214b may be limiting, in view of the diversity of
  upper-atmospheric profiles found at the planetary limb in our 3D
  models.}
\label{fig:TP}
\end{figure*}

\begin{figure}[h!]
\centering \includegraphics[scale=0.35]{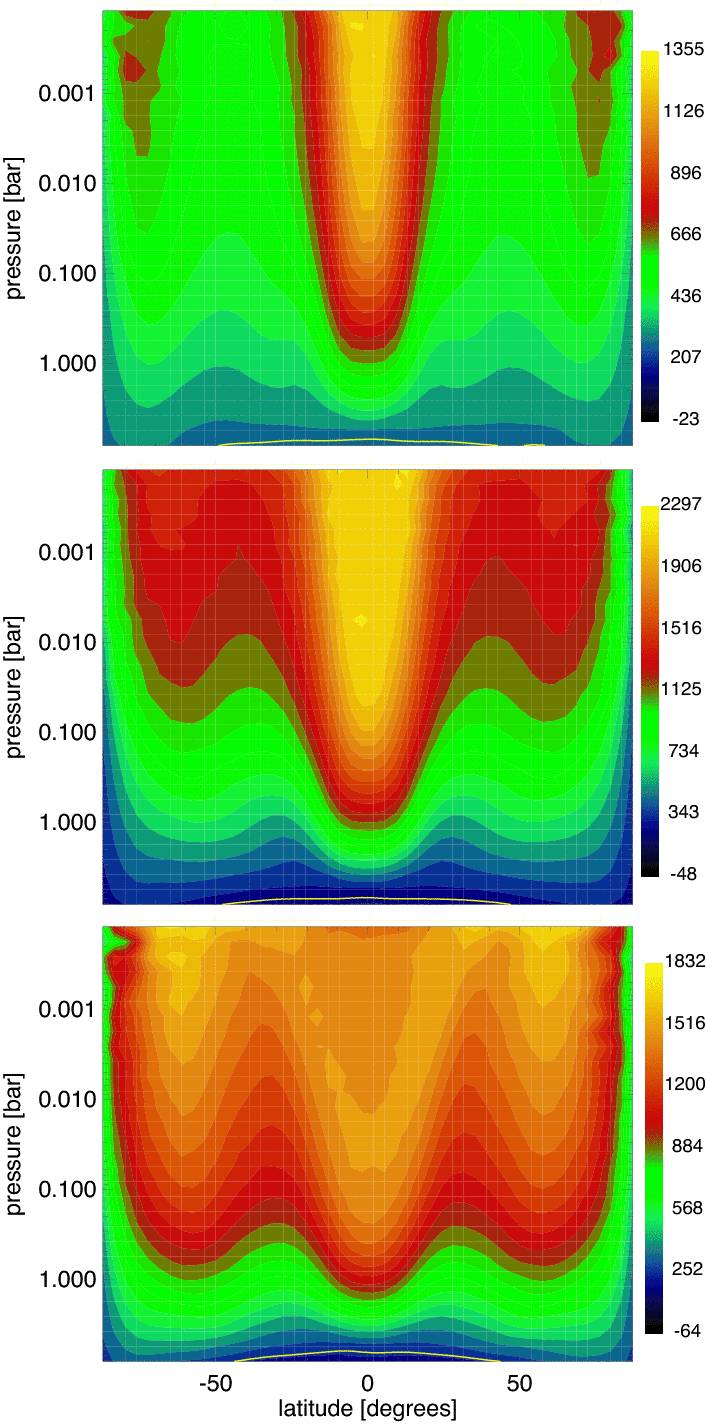}
\caption{Zonal average of the zonal wind (in \ms), as a function of
  latitude and pressure in the atmosphere, for the three models with
  bottom drag (top: Water; middle: SuperSol; bottom: Solar).  A yellow
  line separates the regions of positive (eastward) flow and negative
  (westward) flow.  A prominent super-rotating equatorial jet with
  velocities $\sim 1$-$2$~km~s$^{-1}$ and a transition layer where
  bottom drag is applied are common features of all three models.}
\label{fig:UZ}
\end{figure}

\begin{figure}
\centering \includegraphics[scale=0.4]{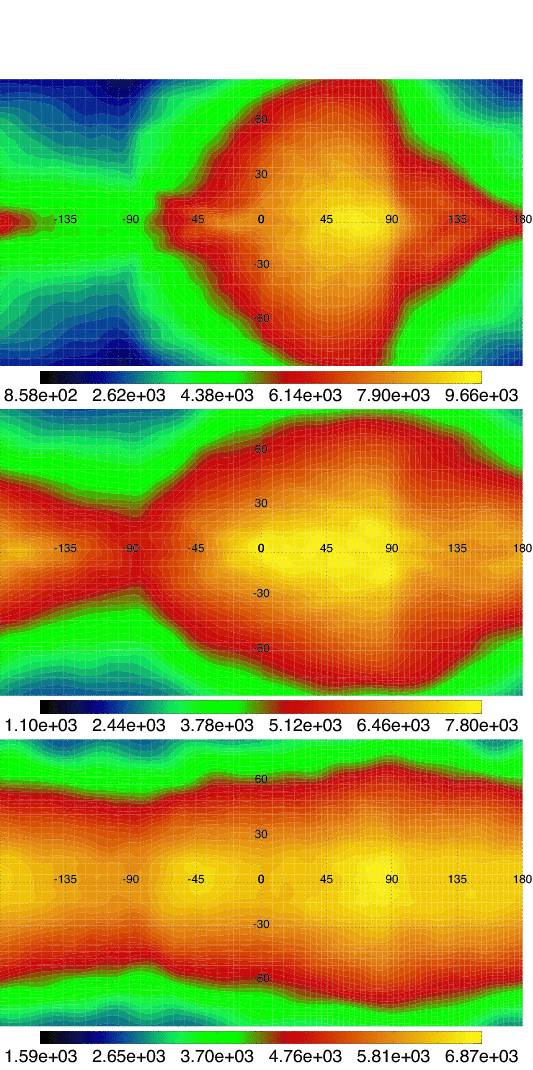}
\caption{Cylindrical maps of the emerging thermal flux in the three
  models with bottom drag (top: Water; middle: SuperSol; bottom:
  Solar). Each map is centered on the substellar point and color-coded
  fluxes are shown in units of W~m$^{-2}$. The degree of eastward
  equatorial heat transport increases significantly from the top to
  the bottom.}
\label{fig:Fmap}
\end{figure}

\begin{figure}
\centering \includegraphics[scale=0.6]{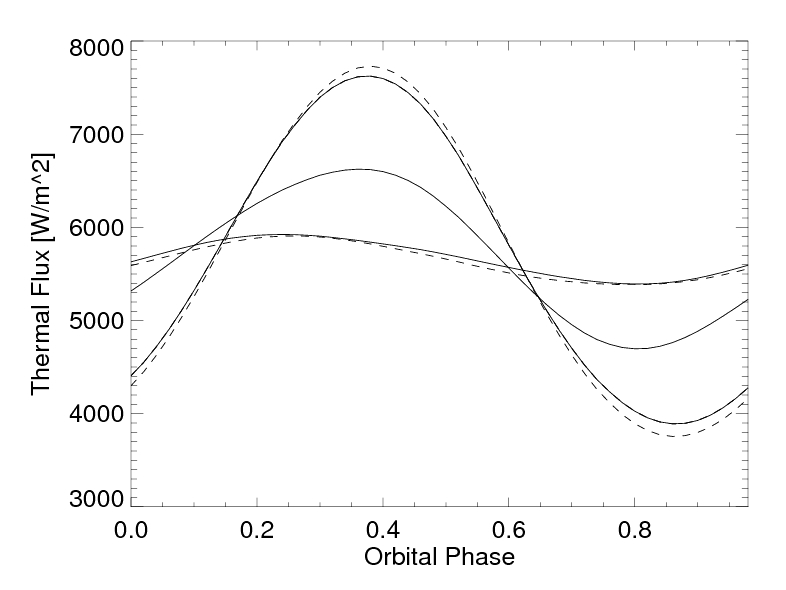}
\caption{Thermal phase curves of the Water (top, at phase 0.4),
  Supersol (middle) and Solar (bottom) models, as seen by a distant
  observer in the planet's equatorial plane.  For each model, phase
  curves are shown in two versions, one with bottom drag (solid line)
  and one without it (dashed line). Transit is at orbital phase 0 and
  secondary eclipse is at orbital phase 0.5. Phase curve amplitudes
  and secondary eclipse depths are systematically reduced from the
  Water, to the SuperSol and finally the Solar model. }
\label{fig:Pcurves}
\end{figure}


\begin{references}

\reference{}
Adams, E. R., Seager, S., \& Elkins-Tanton, L. 2008, ApJ, 673,
1160

\reference{}
Bean, J. L., Miller-Ricci Kempton, E., \& Homeier, D. 2010,
Nature, 468, 669

\reference{}
Bean, J. L. et al. 2011, ApJ submitted, arXiv:1109.0582

\reference{}
Charbonneau, D., et al. 2009, Nature, 462, 891

\reference{}
Cowan, N. B. \& Agol, E. 2011, ApJ 729, 54

\reference{}
Croll, B., Albert, L., Jayawardhana, R., Miller-Ricci Kempton, E.,
Fortney, J. J., Murray, N., \& Neilson, H. 2011, ApJ, 736, 78

\reference{}
Crossfield, I. J. M., Barman, T., \& Hansen, B. M. S. 2011, ApJ,
736, 132

\reference{}
Desert, J.-M., et al. 2011, ApJ, 731, L40+

\reference{}
Guillot, T. 2010, A\&A 520, A27

\reference{}
Henry, G. W. et al. 2011, ApJ submitted, arXiv:1109.2549 


\reference{} Hoskins, B. J. \& Simmons, A. J. 1975,
Quart. J. Roy. Meteo. Soc., 101, 637


\reference{} Forster, de F. P. M., Blackburn, M., Glover, R. \& Shine,
K. P. 2000, Climate Dynamics 16, 833

\reference{} Fortney, J. J., Marley, M. S. \& Barnes, J. W. 2007, ApJ
659, 1661

\reference{}
Lewis, N. K. et al. 2010, ApJ 720, 344 

\reference{}
Lissauer, J. J. et al. 2011, Nature 470, 53


\reference{}
Liu, J., Goldreich, P. M., Stevenson, D. J. 2008, Icar., 653, 664



\reference{}
Menou K. 2011, ApJ submitted, arXiv:1108.3592

\reference{}
Menou, K. \& Rauscher, E. 2009,  ApJ 700, 887

\reference{}
Miller-Ricci, E., Seager, S., \& Sasselov, D. 2009, ApJ, 690, 1056

\reference{}
Miller-Ricci, E. \& Fortney, J. J. 2010, ApJ 716, L74



\reference{} Miller-Ricci Kempton, E., Zahnle, K. \& Fortney,
J. J. 2011, arXiv:1104.5477

\reference{}
Nettelmann, N., Fortney, J. J., Kramm, U., \& Redmer, R. 2011,
ApJ, 733, 2

\reference{}
Perna, R., Menou, K. \& Rauscher, E., 2010, ApJ 719, 1421

\reference{}
Rauscher, E. \& Menou, K. 2010, ApJ 714, 1334

\reference{}
Rauscher, E. \& Menou, K. 2011a, ApJ submitted, arXiv:1105.2321


\reference{} Rauscher, E. \& Menou, K. 2011b, to be submitted to ApJ

\reference{} 
Rogers, L. A. \& Seager, S. 2010, ApJ 716, 1208

\reference{} 
Schneider, T. \& Liu, J. 2009, Journal of Atmospheric Sciences, 66, 579

\reference{} 
Showman, A. P., Y-K. Cho, J., Menou, K. 2010, in
``Exoplanets'', Space Science Series of the University of Arizona
Press (Tucson, AZ); Ed. S. Seager; arXiv:0911.3170.



\reference{} 
Showman, A. P. \& Polvani, L. M. 2011, ApJ 738, 71

\reference{} Thrastarson, H. T. \& Cho, J. Y-K. 2011, ApJ 729, 117

\end{references}
\end{document}